\documentclass[11pt,a4paper]{article}
\usepackage{amsmath,amssymb,amsthm,amscd}
\usepackage{epsfig,comment}
\usepackage{graphicx,caption,subcaption}
\usepackage{yfonts,mathrsfs}
\usepackage{graphics,graphicx,epsfig}
\usepackage{amsfonts,psfrag,verbatim,color}
\def\a{\alpha}

\def\r{\rho}
\def\s{\sigma}
\def\t{\tau}
\def\m{\mu}
\def\n{\nu}
\def\k{\kappa}
\def\th{\theta}
\def\g{\gamma}\def\G{\Gamma}
\def\L{\Lambda}\def\l{V}
\def\D{\Delta}
\def\la{\langle}
\def\ra{\rangle}
\def\o{\omega}\def\O{\Omega}
\def\d{\delta}
\def\p{\partial}

\def\oxthree{{\cal O}(x^3) }

\def\half{\textstyle{\frac{1}{2}}}

\def\bdoc{\begin{document}}
\def\edoc{\end{document}}
\def\bea{\begin{equation}}
\def\eea{\end{equation}}

\def\beq{\begin{eqnarray}}
\def\eeq{\end{eqnarray}}
\def\ben{\begin{enumerate}}
\def\een{\end{enumerate}}
\def\la{\langle}
\def\ra{\rangle}
\def\a{\alpha}
\def\g{\gamma}\def\G{\Gamma}
\def\d{\delta}\def\D{\Delta}
\def\e{\epsilon}
\def\z{\zeta}

\def\th{\theta}
\def\k{\kappa}
\def\l{\lambda}
\def\m{\mu}
\def\n{\nu}
\def\o{\omega}
\def\p{\pi}
\def\r{\rho}
\def\s{\sigma}
\def\t{\tau}
\def\L{{\Lambda}}
\def\S{\Sigma }
\def\gsim{\; \raisebox{-.8ex}{$\stackrel{\textstyle >}{\sim}$}\;}
\def\lsim{\; \raisebox{-.8ex}{$\stackrel{\textstyle <}{\sim}$}\;}
\def\gtrsim{\gsim}
\def\lessim{\lsim}
\def\loc{{\rm local}}
\def\vm{v_{\rm max}}
\def\bh{\bar{h}}
\def\del{\partial}
\def\nab{\nabla}
\def\half{{\textstyle{\frac{1}{2}}}}
\def\fourth{{\textstyle{\frac{1}{4}}}}
\def\h{\mathscr{H}}
\def\bD{{\bf D}}
\def\bE{{\bf E}}
\def\bF{{\bf F}}
\def\bB{{\bf B}}
\def\bP{{\bf P}}
\def\bV{{\bf v}}
\def\bv{{\bf v}}
\def\bx{{\bf x}}
\def\by{{\bf y}}
\def\bz{{\bf z}}
\def\ba{{\bf a}}
\def\bd{{\bf d}}
\def\bs{{\bf s}}
\def\bn{{\bf n}}
\def\bp{{\bf p}}

\def\O{\Omega}

\def\br{{\bf r}}
\def\bnab{{\bf \nab}}

\def\tE{\tilde{E}}
\def\tL{\tilde{L}}
\def\Horava{Ho\v{r}ava }

\def\oxtwo{\mathscr{O}\left(x^2\right)}
\def\oxthree{\mathscr{O}\left(x^3\right)}
\def\oxfour{\mathscr{O}\left(x^4\right)}
\def\oxfive{\mathscr{O}\left(x^5\right)}

\def\ph{\phantom}

\begin{document}

\title{Boundary Conservation from Bulk Symmetry \footnote{ Essay written for the Gravity Research Foundation 2018 Awards for Essays on Gravitation}}
\author{C. Fairoos \footnote{fairoos.c@iitgn.ac.in, corresponding author}, Avirup Ghosh\footnote{avirup.ghosh@iitgn.ac.in} and Sudipta Sarkar\footnote{sudiptas@iitgn.ac.in}\\
\small{Indian Institute of Technology, Gandhinagar, 382355, Gujarat , India}\\
}
\maketitle


\begin{abstract}
The evolution of the black hole horizon can be effectively captured by a fictitious membrane fluid living on the stretched horizon. We show that the dynamics of this boundary matter arises from the invariance of the bulk action under local symmetries in the presence of the inner boundary. If general covariance is broken in a semi-classical treatment of a quantum field near a black hole horizon, we argue that it can be restored by the inclusion of a quantum flux into the membrane conservation equation which is exactly equal to the Hawking flux. 
\end{abstract}

Black hole event horizon divides the space-time into two causally disconnected regions. This causal separation is an advantage as it allows us to practice low energy physics without worrying about the nature of space-time near the central singularity. This is reminiscent of the Wilsonian decoupling of UV from IR in a well-behaved quantum field theory. The unobserved UV degrees of freedom leave their marks in the renormalization group flow of the couplings. We hope for similar signatures of the unobserved information inside the black holes as an effective quantity for the outside observer. A novel description of such an approach is the construction of the membrane paradigm \cite{Price:1986yy, Parikh:1997ma}. Suppose we live in a universe which has an inner boundary, akin to the black hole event horizon. Unlike the asymptotic boundary, the inner boundary is not a real boundary of the space-time. Some observers can venture into the inside from outside. Nevertheless, the outside physics is causally independent of the inside. Then, we should be able to derive a classical equation of motion without imposing any special boundary condition at the inner boundary. For example, consider the action of a Maxwell field sourced by a charged scalar field,
\begin{equation}\label{action_1}
S = \int_{M}  d^4x \sqrt{-g} \Big[ -\frac{1}{4} F_{\mu\nu}F^{\mu\nu} - m^2 \phi^* \phi + (\mathcal{D}_\mu \phi)^*(\mathcal{D}^\mu\phi)\Big],
\end{equation}
where $ \mathcal{D}_\mu \phi= \nabla_\mu \phi - i e A_\mu \phi $ is the gauge covariant derivative of the field $\phi$. The variation of the action leads to Maxwell's equations provided we impose suitable boundary conditions at infinity to kill the boundary term resulting from the variation. Suppose the space-time contains an inner boundary besides the one at the infinity. Then the variational principle becomes ill-defined due to the presence of the boundary term which does not vanish at the inner boundary, without any special boundary condition. So, the stationarity of the on-shell action requires that we add an appropriate additional boundary term in the variation which cancels the offending one. The additional term is interpreted as due to a ``fictitious" membrane matter living on the inner boundary. The membrane is therefore described by a boundary current density $j^{\mu} = F^{\mu \nu } n_\nu$ and boundary scalar fields $\Phi^* = \left(\mathcal{D}^\m \phi^*\right)n_\m ; \Phi= \left(\mathcal{D}^\m \phi\right)n_\m$, where $n_\nu$ is the (outward) normal of the boundary. We have assumed the inner boundary to be time-like. For black hole event horizon, we consider the corresponding time-like stretched surface. The existence of the membrane matter is a classic example of matter replacing the role of the boundary condition.\\

But the question remains, in what sense this fictitious membrane matter captures the effective dynamics of the black hole? In the traditional construction of membrane paradigm, this is answered by direct derivation of the equation \cite{Parikh:1997ma}
\beq
D_\n j^\n  = J^\n n_\n,
\eeq
 where $J^\n = ie \left(\phi^* \mathcal{D}^\n \phi - \phi \mathcal{D}^\n \phi^* \right)$ is the current density of the bulk charged matter and $D_\m$ is the covariant derivative compatible with the induced metric on the inner boundary. This equation is interpreted as the conservation equation of the membrane, where the flow of bulk matter balances the divergence of the boundary current density. It shows that the membrane is dynamic and responsive to changes due to the matter falling into the inner boundary. The fictitious membrane is therefore endowed with life, imitating the dynamics of matter lost into the boundary.\\
 
In this article, we provide a novel understanding of this equation from the symmetry of the bulk. We show that the requirement of local gauge invariance in the presence of an inner boundary leads to this conservation equation of the boundary matter. The bulk symmetry enforces conservation in the boundary. \\

We consider the following local gauge transformations of the fields,
\begin{equation}
\phi' =\mathrm{e}^{ie \lambda(x)} \phi ;\qquad A_\m' = A_\m + \nabla_\m \lambda(x).
\end{equation}
The variation of the action under this gauge transformation, on shell, takes the form of a boundary term. All the contributions from the outer boundary vanish provided the gauge function $\lambda (x)$ vanishes at infinity. But, at the inner boundary, we have a non vanishing contribution:
\begin{equation}
\d_\lambda S = \int d^3x \sqrt{-h} \Big[ D_\n\left(F^{\m\n}n_\m \right) \lambda(x) - J^\m n_\m \lambda(x) \Big].
\end{equation}
If we demand that the bulk action be invariant under the local gauge transformation, even in the presence of an inner boundary, we need set the integrand to zero and this will lead to the desired conservation equation $D_\n j^\n  = J^\n n_\n$  of the dynamic membrane matter.  The conservation arises from the symmetry of the bulk action in the presence of the boundary.\\
 
 The same result can be obtained for the gravitational membrane also. Consider the action of a scalar field coupled to gravity in a curved space-time with inner boundary, a black hole stretched horizon,
 
 \begin{equation}
S = \frac{1}{2} \int_M d^4x \sqrt{-g}\,\Big(R - g^{\m\n} \phi_{,\m}\phi_{,\n}\Big)+ \int_{\partial M} d^3x \sqrt{ h} \,K.
\end{equation} 
 
We have added the Gibbons-Hawking-York's boundary terms containing the trace of the extrinsic curvature $K$ of the boundary, which has an intrinsic metric $h_{\m \n}$. The membrane matter for this action is described by a boundary stress tensor $t_{\mu \nu} = K h_{\m\n} - K_{\m\n}$ and boundary field $\Phi = n^\m\nabla_\m\phi$.\\

 We demand the invariance of the action under the infinitesimal diffeomorphism: $x'^\m = x^\m + \xi^\m$. In the absence of the inner boundary, action is indeed diffeomorphism invariant as the vector field $\xi^\m$ is set to vanish at infinity. But, once the inner boundary is present and we do not set any condition on $\xi$ except that it is tangent to the inner boundary, we will have a non vanishing term at the inner boundary given by,
 
 \begin{equation}
\delta_{\xi} S = -\int d^3x \sqrt{-h}\Big\{ \xi^\n D^\m\,t_{\mu \nu}   - n^\m \xi^\n\left(\partial_\m\phi  \, \partial_\n \phi \right)  \Big\}.
\end{equation}

The requirement of the diffeomorphism invariance of the action: $ \delta_{\xi} S = 0$ gives,
\begin{equation}
D^\m t_{\m \n} =  - n^\m \partial_\m\phi\, \partial_\n \phi  = - h^\alpha_\n T_{\alpha \m} n^\m.
\end{equation}

The equation has the obvious interpretation as the conservation equation for the gravitational membrane. It captures the response of the membrane when the matter flux enters into the boundary.\\

The local gauge invariance of the theory implies the existence of redundant degrees of freedom. In general, such an invariance does not lead to any conservation law. In fact, the parameters of the gauge transformations, like the function $\lambda(x)$ or the diffeomorphism generating vector field $\xi^\mu$, are set to be trivial at the boundary. The presence of the inner boundary affects this construction because the gauge parameter cannot be set to vanish on the inner surface. As a result, the local invariance of the bulk theory requires an extra condition, namely the conservation law of the boundary matter. The conservation law makes the membrane matter dynamic,  providing an effective description of the interior. In case of black hole horizon, the limit of the stretched horizon to the actual horizon is taken, and various projections of this conservation law give the energy and momentum evolution equation of the membrane. These equations are in the form of equations of viscous fluid dynamics with appropriate bulk and shear viscosities \cite{Price:1986yy, Parikh:1997ma}. All these can be seen as the requirement of the diffeomorphism invariance of the bulk, which enforces the boundary conservation.\\

Let us compare our result with the construction based on horizon constraints developed in the works \cite{Carlip:1999cy, Carlip:2002be, Carlip:2004mn}. The presence of the stretched horizon as an inner boundary has the effect of adding a central extension to the algebra of diffeomorphisms. The central extension provides the asymptotic behavior of the density of states producing the Bekenstein entropy of the horizon. As in our construction, the imposition of horizon boundary conditions alters the physical content of the theory and create dynamical degrees of freedom from the pure gauge. The conservation law of the membrane at the boundary, resulting from the local symmetry of the bulk is a vivid illustration of this idea. \\

The membrane conservation law arising from local gauge symmetry of the classical Lagrangian takes into account only the classical flux of matter into the boundary. Consideration of the quantum effects may add another interesting twist. It is suggested \cite{Robinson:2005pd, Iso:2006wa} that the regularity of the modes on the horizon requires the effective theory, for the behavior of fields in the region outside the horizon, to be chiral. This leads to the breakdown of the general covariance. The flux required to cancel the gravitational anomaly at the horizon has a form equivalent to blackbody radiation with Hawking temperature. Therefore, in our picture, since the diffeomorphism invariance does not hold true near the horizon, the gravitational membrane conservation equation also breaks down by quantum effects and we have a tentative equation $ D^\m t_{\m \n} + h^\alpha_\n T_{\alpha \m} n^\m = {\cal O}{(\hbar)}$. If we require that the diffeomorphism invariance holds true even in the semi-classical level, we need another quantum flux to compensate the right-hand side. This flux must be same as the flux $J_H$ of blackbody radiation at Hawking temperature $T_H$. Therefore, we would have a quantum conservation law:

 \begin{equation}
D^\m t_{\m \n} + h^\alpha_\n T_{\alpha \m} n^\m  + J_{H} = 0
\end{equation}

 This semi-classical conservation equation captures the response of the membrane due to both classical matter flux as well as quantum Hawking flux. Note that this would be an exact equation within semi-classical approximation containing all information about the back reaction effects. The evolution of this quantum membrane could be useful to find correlations in the outgoing Hawking spectrum. \\
 
In our construction, the effective degrees of freedom and the laws of evolution of black hole have their origin in the bulk symmetry. It is then natural to expect that the black hole entropy of Killing horizons will be closely related to the Noether charge of Killing isometry as shown in the Wald's formalism of the first law \cite{Wald:1993nt}. The main ingredients in all these are the symmetries of the bulk which controls the physics of the boundary. One wonders, in the spirit of the holographic principle, if this can be reversed?

\section*{Acknowledgement}
We thank Maulik Parikh for discussion and suggestions. AG is supported by SERB, Government of India through the National Post Doctoral Fellowship grant
(PDF/2017/000533). SS is supported by the SERB Fast Track Scheme for Young Scientists (YSS/2015/001346).



\end{document}